\documentclass[a4paper]{jpconf}
\usepackage{graphicx}
\usepackage{amsmath,amssymb}

\def\be{\begin{equation}}
\def\ee{\end{equation}}
\def\beq{\begin{eqnarray}}
\def\eeq{\end{eqnarray}}

\begin{document}
\title{Perturbations in loop quantum cosmology}

\author{W Nelson\footnote{This talk is based on the joint work of reference~\cite{AAN}}, I Agullo  and A Ashtekar.}

\address{Institute for Gravitation and the Cosmos, The Pennsylvanian State University, USA.}

\ead{nelson@gravity.psu.edu}

\begin{abstract}
The era of precision cosmology has allowed us to accurately determine many important cosmological parameters, in particular via the CMB.
Confronting Loop Quantum Cosmology with these observations provides us with a powerful test of the theory. For this to be possible we
need a detailed understanding of the generation and evolution of inhomogeneous perturbations during the early, Quantum Gravity, phase
of the universe. Here we describe how Loop Quantum Cosmology provides a completion of the inflationary paradigm, that is consistent
with the observed power spectra of the CMB.
% dynamics of the background alters the power spectra of
%the CMB.
\end{abstract}

\section{Open issues with inflation}
The standard picture we have of early universe cosmology is that of inflation. This beautiful idea contains very few basic assumptions,
yet it produces, with spectacular agreement, the power spectra of density fluctuations observed in the Cosmic Microwave Background (CMB).
Nevertheless, inflation remains a paradigm in search of a model and there are several key questions that remain unanswered. Broadly
speaking, these can be split into two categories: difficulties facing the particles physics interpretation of inflation, and those we expect
to be related to quantum gravity. Examples of each type are given in Table~(\ref{tab:1}).

Inflation is a period of quasi-de Sitter expansion and typically 
the assumption is that slow-roll inflation began with the quantum state describing perturbations in its `natural' 
vacuum state: the Bunch-Davies vacuum. A priori this is an odd assumption, since it says that the 
quantum state is tuned to the subsequent (quasi-de Sitter) evolution of the geometry. Essentially this implies that 
the pre-inflationary dynamics of the universe and the `true' initial state conspired in such a way as to ensure that
we arrived at the onset of slow-roll inflation with no particles present (relative to the Bunch-Davies vacuum). 
The intuition behind this assumption was that even if there were particles present at the onset of inflation,
the exponential expansion would rapidly dilute them and hence their consequences can safely be ignored. This
intuition misses the important fact that quantum fields in a dynamical space-time experience {\it both}
spontaneous and simulated creation of particles~\cite{spon_emm}. The latter effect actually compensates for the 
exponential growth of the volume in such a way that the particle number density remains approximately constant.

In classical general relativity inflation is inevitably preceded by the Big-Bang and 
the only natural place to give initial conditions is at this singularity. An important open question then is:
can one find a quantum gravity completion to the inflationary paradigm? To be viably, such a completion should be 
non-singular and agree with current observations. It would also open up the exciting possibility that we can
directly observe the pre-inflationary universe. If it were possible to see observational consequences of the quantum
state at the onset of slow-roll, we would be able to probe the dynamics of the quantum gravity era of the universe.

\begin{table}
\begin{center}
\begin{tabular}{|c|c|}
\hline
Particle physics issues & Quantum gravity issues \\
\hline
What is the inflaton? & What are the initial conditions? \\
Why is the potential flat? & Is there a singularity? \\
How does it couple to the standard model? & Why is perturbation theory valid? \\
What interactions are present? & What happens when the frequencies \\
 &  become transplanckian? \\
\hline
\end{tabular}
\caption{\label{tab:1} Some examples of the unresolved issues facing inflation.}
\end{center}
\end{table}

\section{Loop Quantum Cosmology}
Loop Quantum Gravity is a particularly well developed approach to quantising gravity~\cite{LQG}. It is
a Hamiltonian quantisation that maintains
the fundamental relationship between geometry and gravity and is fully non-perturbative. However a rigorous
understanding of the dynamics of the theory are still lacking (see Madhavan Varadarajan's talk in this session). One useful way to
make progress is to consider simplified (truncated) systems of the full theory, which, on the one hand,
can be completely understood, and on the other are physically interesting. This approach has been applied with great 
success to the study of black-holes (see Fernando Barbero's talk in this session) and graviton propagators
(see Carlo Rovelli's talk in this session) within Loop Quantum Gravity. One can also consider
the truncation of full general relativity to homogeneous systems and then use Loop Quantum Gravity
techniques to quantise these. This leads to Loop Quantum Cosmology (LQC)\cite{LQC}, which has turned out
to be very successful at answering many of the difficulties facing classical cosmology. In particular,
it has been shown in detail how the classical singularity is replaced by a `Big-Bounce' and
how the late time, low energy, limit reproduces the standard expectations of general relativity.
It has also been shown that the probability of having a sufficiently long phase of (single scalar field driven) slow-roll inflation, within LQC, is
very close to one~\cite{Probability}.

However in order to make predictions about LQC effects on the CMB, one first has to extend the underlying
approach to include (peturbative) inhomogeneities. There have been several promising attempts to include
such inhomogeneities (see Jakub Mielczarek's talk in this session) based on consistent alterations of 
the homogeneous formulation of LQC. Here we describe a systematic approach to extending
the underlying formulation.

\section{LQC and perturbations}
The first step is to find a suitable truncation of the phase-space of classical general relativity
that allows for cosmologies with perturbative inhomogeneities. Since we work in the Hamiltonian theory,
we restrict our attention to cosmologies whose spatial slices are (flat) tori. One can then 
show that the full phase-space decomposes into homogeneous and {\it purely} inhomogeneous parts,
\be
\Gamma_{\rm Full} = \Gamma_{\rm H} \times \Gamma_{\rm IH}~.
\ee
The canonical variables of the homogeneous phase-space ($\Gamma_{\rm H}$) are $\left( a,\pi_a,\phi,\pi_\phi\right)$, 
i.e.\ the scale factor $a$ and the scalar field $\phi$, and their conjugate momenta. The
canonical variables of the inhomogeneous phase-space ($\Gamma_{\rm IH}$) are the corresponding perturbations, $\left( h_{ab}\left(x\right),
  \pi^{ab}\left(x\right),\varphi\left(x\right),\pi_\varphi\left(x\right)\right)$, all of which are purely inhomogeneous.

All the canonical structures of the phase-space decompose in this way, in particular the symplectic
structure and the Possion brackets factor. One can now consider the inhomogeneous fields as perturbations,
expand the constraints and find gauge invariant degrees of freedom~\cite{langlois}.
In particular the gauge invariant scalar modes of the perturbations satisfy the constraint,
\be\label{eq:Ham_class}
{\cal C}= \frac{\pi_\phi^2}{2l^3} + \frac{m^2V^2}{2l^3} - \frac{3}{8\pi G l^3} b^2 V^2
 +\int {\rm d}^3 k\left( \frac{1}{2} P_k^2 + f\left( V,b,\phi,\pi_\phi; k\right) Q^2_k\right) \approx 0~,
\ee
where $V\sim a^3$ and $b\sim \pi_a/a^2$ are canonically conjugate background variables, $l$ is
the coordinate size of the 3-torus and $\left( P_k, Q_k \right)$ are the gauge invariant, scalar
degree of freedom (related to the Mukhanov variable).
The important point is that gauge invariant scalar (and tensor) perturbations behave exactly as
test scalar fields with a time dependent mass $f\left( V,b,\phi,\pi_\phi; k\right)$.

In~\cite{QFT_QBG} it was shown how to quantise such a system, and we schematically sketch this procedure
here. Writing the background and perturbation parts of Eq.~(\ref{eq:Ham_class}) as,
\be
 {\cal C} = \frac{\pi_\phi^2}{2 l^3}+H^2_0 + \int {\rm d}^3 k H_{\tau,k}~,
\ee
where $H_0$ is the Hamiltonian for the background geometry (and the potential term)
and $H_{\tau,k}$ is the (time dependent) Hamiltonian 
for the perturbations. These are then promoted to operators and we deparameterise with respect
to the scalar field $\phi$, to find,
\be\label{eq:Ham_quant}
 -i \hbar \partial_\phi \Psi = \left( \widehat{H_0^2} + \int {\rm d}^3 k\widehat{H}_{\tau,k} \right)^{1/2} \Psi
\approx \left(\widehat{H}_0  + \widehat{H}_0^{-1/2}\left( \int{\rm d}^3 k
\widehat{H}_{\tau,k}\right)  \widehat{H}_0^{-1/2} \right) \Psi~,
\ee
where $\widehat{H_0^2} = \hbar^2 \widehat{\Theta}_0$ is the operator that governs the LQC evolution of the background~\cite{LQC},
and the right hand side has been approximated via a series expansion.
%One now approximates the right hand side of this operator equation via a series expansion,
%\be\label{eq:Ham_quant}
% -i\hbar \partial_\phi \Psi \approx \left(\widehat{H}_0  + \widehat{H}_0^{-1/2}\left( \int{\rm d}^3 k
%\widehat{H}_{\tau,k}\right)  \widehat{H}_0^{-1/2} \right) \Psi~.
%\ee
This is the only approximation that is made in the procedure and it can be viewed as a
test-field approximation. Essentially, one restrictions ones attention to those states in which
the $\widehat{H}_0$ is dominant and $\widehat{H}_{\tau,k}$ does not alter the background
dynamics\footnote{There are many important issues to do with ensuring that the integral over
$k$ results in a well defined operator that are not being described here. See~\cite{AAN} for 
details.}.
The second term on the right hand side of Eq.~(\ref{eq:Ham_quant})
% The latter operator,
%\be
% \widehat{H}_0^{-1/2} \left(\int{\rm d}^3 k
%\widehat{H}_{\tau,k}\right)  \widehat{H}_0^{-1/2}~,
%\ee
is precisely the Hamiltonian for the perturbations associated to the relational time defined by $\phi$.

Finally, one decomposes the wave-function as a tensor product of back-ground and perturbation pieces,
\be
 \Psi \left( V, Q_k,\phi\right) = \Psi_0\left( V,\phi\right) \otimes \psi\left( Q_k, \phi\right)~,
\ee
and considers states for which $\Psi_0$ is sharply peaked with respect to some particular classical background
geometry at late times i.e.\ a semi-classical background geometry. Taking expectation values of Eq.~(\ref{eq:Ham_quant})
with respect to $\Psi_0$, one arrives the standard Hamiltonian for test quantum fields in a curved space-time,
with the background scale factor $a\left(\phi\right)$ replaced by $\langle \Psi_0\big|\hat{a}\big|\Psi_0
\rangle \big|_{\phi}$.

\section{Results and conclusions}
In the previous section we briefly sketched how one can extend the formulation of LQC to include (perturbative)
inhomogeneities. With this in hand, one can now look for consequences of the pre-inflationary era and in particular,
we can look for deviations from the standard initial conditions for inflation. The standard approach in inflation is to assume the
existence of a scalar field with a suitable potential, give oneself some initial conditions for the quantum
state of the perturbations and hence calculate the late time power spectra. Here we will take exactly the same
approach: we specify initial conditions at the Big-Bounce
and calculate the resulting power-spectra (here we concentrate only on tensor perturbations).
Since, at the bounce, the wavelength of all the observable modes is much smaller than the curvature scale, 
we take the initial state (for these modes) to be the Minkowski vacuum (see~\cite{AAN} for a discussion on this point).

\begin{figure}
\begin{center}
\includegraphics[scale=0.65]{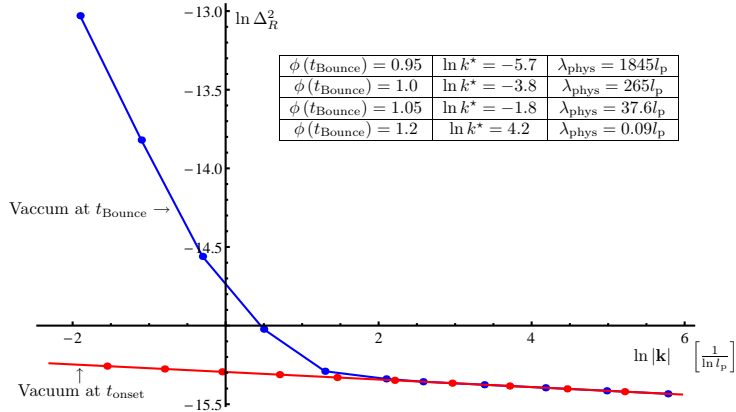}
\caption{\label{fig:1} The power-spectra for tensor perturbations at the end of slow-roll inflation, given an initial 
vacuum at the bounce. We have set $a\left(t_{\rm bounce}\right) =1$ and hence the physical
scale at (for example) decoupling depend on the amount of inflation that has occurred. Note in particular the agreement
with the predictions of standard inflation provided $\phi\left(t_{\rm bounce}\right) > 1.2$.}
\end{center}
\end{figure}

Figure~(\ref{fig:1}) shows the resulting late time (dimensionless) power spectra, $\Delta^2_R$, defined as,
\be
 \delta\left({\bf k} + {\bf k}'\right) \frac{\Delta^2_R\left(k\right)}{4\pi |{\bf k}|^3} =
\langle 0| \hat{R}_{\bf k} \hat{R}_{\bf k'}|0\rangle ~,
\ee
where the variable $R_{\bf k}$ is related to the the Mukhanov variable $Q_{\bf k}$ by $R_{\bf k} = \frac{\dot{\phi}}{H} Q_{\bf k}$,
and represents the gravitational potential on constant $\phi$ hyper-surfaces.

The important result here is that provided the background scalar field, $\phi$, at the bounce is large enough, i.e.\ 
$\phi\left(t_{\rm bounce}\right) \geq 1.2$, the power spectra is entirely consistent with that of standard 
inflation. Hence we have a completion of the inflationary paradigm, including the quantum gravity era, which agree with the
observations.  Note that there remains a (small but important) window ($0.95 < \phi\left(t_{\rm bounce}\right)< 1.2$) in which we may hope to see
corrections to the largest scales (smallest $k$) of the observed power spectra, and hence directly
observe pre-inflationary (i.e.\ quantum gravity) physics.

\subsection{Acknowledgments}
This work was supported in part by the NSF grants PHY0854743 and PHY1068743.

\section{References}
%%%%%%%%%%%%%%%%%%%%%%%%%%%%%%%%%%%%%%%%%%%


\begin{thebibliography}{9}
\bibitem{AAN} I. Agullo, A. Ashtekar, W. Nelson,~ {\it In prep.}.
\bibitem{spon_emm} I. Agullo, L. Parker,~ Phys. Rev. D {/bf 83} 063526 (2011).
\bibitem{LQG} A. Ashtekar, M. Bojowald, J. Lewandowski,~ Adv. Theor. Math. Phys. {\bf 7} 233-268 (2003), A. Ashtekar, J. Lewandowski,~ Class.
Quant. Grav. {\bf 21} R53 (2004)~.
\bibitem{LQC} A. Ashtekar, P. Singh,~ Class. Quant. Grav. {\bf 28} 213001 (2011), M. Bojowald,~ Living Review {\bf 11} (2008) (cited on 20 March 2012)~.
\item Big-bounce refs.
\bibitem {Probability} C. Germani, W. Nelson, M. Sakellariadou,~ Phys. Rev. D {\bf 76} 043529 (2007),
                       A. Ashtekar, D. Sloan,~ Gen. Rel. Grav. {\bf 43} 3619-3656 (2011),
                       A. Corichi, A. Karami,~ Phys. Rev. D {\bf 83} 104006 (2011).
\bibitem{langlois} David Langlois,~ Class. Quantum. Grav. {\bf 11} 387-407 (1994)~.
\bibitem{QFT_QBG} A. Ashtekar, W. Kaminski, J. Lewandowski,~ Phys. Rev. D. {\bf 79} 064030 (2009).
\end{thebibliography}
\end{document}